\documentclass[]{aastex631}
\usepackage{multirow}

\usepackage{tabularx}
\usepackage{graphicx} 
\usepackage{times}
\usepackage{url}
\usepackage{ctable}
\usepackage{enumitem}
\usepackage{float}
\usepackage{xcolor}
\usepackage{comment}
\usepackage{hyperref}

\usepackage{lineno}
\linenumbers

\begin{document} 

\nolinenumbers

\title{Using gravitational waves to distinguish between neutron stars and black holes in compact binary mergers}

\author{Stephanie M. Brown}
\affiliation{Albert-Einstein-Institut, Max-Planck-Institut f{\"u}r Gravitationsphysik,\\ 
Callinstra{\ss}e 38, 30167 Hannover, Germany,}
\affiliation{Leibniz Universit{\"a}t Hannover, 30167 Hannover, Germany,}

\author{Collin D. Capano}
\affiliation{Albert-Einstein-Institut, Max-Planck-Institut f{\"u}r Gravitationsphysik,\\
Callinstra{\ss}e 38, 30167 Hannover, Germany,}
\affiliation{Leibniz Universit{\"a}t Hannover, 30167 Hannover, Germany,}

\author{Badri Krishnan}
\affiliation{Albert-Einstein-Institut, Max-Planck-Institut f{\"u}r Gravitationsphysik,\\ Callinstra{\ss}e 38, 30167 Hannover, Germany,}
\affiliation{Leibniz Universit{\"a}t Hannover, 30167 Hannover, Germany,}
\affiliation{Institute for Mathematics, Astrophysics and Particle Physics, Radboud University\\ 
Heyendaalseweg 135, 6525 AJ Nijmegen, The Netherlands}

\begin{abstract}
In August 2017, the first detection of a binary neutron star merger, GW170817, made it possible to study neutron stars in compact binary systems using gravitational waves. Despite being the loudest gravitational wave event detected to date (in terms of signal-to-noise ratio), it was not possible to unequivocally determine that GW170817 was caused by the merger of two neutron stars instead of two black holes from the gravitational-wave data alone. That distinction was primarily due to the accompanying electromagnetic counterpart. This raises the question: under what circumstances can gravitational-wave data alone, in the absence of an electromagnetic signal, be used to distinguish between different types of mergers? Here, we study whether a neutron star–black hole binary merger can be distinguished from a binary black hole merger using gravitational-wave data alone. We build on earlier results using chiral effective field theory to explore whether the data from LIGO and Virgo, LIGO A+, LIGO Voyager, the Einstein Telescope, or Cosmic Explorer could lead to such a distinction. The results suggest that the present LIGO–Virgo detector network will most likely be unable to distinguish between these systems even with the planned near-term upgrades. However, given an event with favorable parameters, third-generation instruments such as Cosmic Explorer will be capable of making this distinction. This result further strengthens the science case for third-generation detectors.
\end{abstract}

\section{Introduction}

Neutron stars are unique laboratories for studying ultra-dense, relativistic matter. Multimessenger observations of neutron star mergers provide unique opportunities to extract relevant physical information (such as compactness) from them. Measurements of neutron star compactness and radii are vital to constraining the equation of state of ultra-dense matter \cite{Lattimer_2001}. In addition to gravitational wave observations of the neutron star mergers GW170817 and GW190425 \cite{LVC_2017,Abbott_2020}, X-ray observations of accreting neutron stars \cite{Ozel_2016,Watts_2016} have placed constraints on neutron star mass and radii. Of these, the recent results from NICER are especially promising \cite{Bogdanov_2019a,Bogdanov_2019b,Raaijmakers_2020}.

The observation of GW170817 and its electromagnetic counterpart led to several important advances. The detection of the electromagnetic counterpart was possible because the LIGO–Virgo observation constrained the sky location of the event to 28 deg$^2$. It was the detection of gamma-ray burst GRB170817A 1.7 seconds after GW170817 that provided the initial evidence that this event contained neutron star matter. Transient electromagnetic follow-ups \cite{Soares_Santos_2017,Cantiello_2018} further supported the neutron star hypothesis and provided more information about the binary. The combination of electromagnetic and gravitational wave observations led to new constraints on neutron star physics. For instance, analyses of GW170817 placed upper limits on the radius of a $1.4\,\mathrm{M}_\odot$ neutron star: $11.0^{+0.9}_{-0.6}$ km \cite{capano_2020}, $12.2^{+1.0}_{-0.8}\pm0.2$ km \cite{Radice_2019}, $10.8^{+2.4}_{-1.9}$ km \cite{De_2018}. The LIGO–Virgo Collaboration constrained the radii of the two components of GW170817 ($M_{1} = (1.36,1.58) M_{\odot}$, $M_{2} = (1.18,1.36) M_{\odot}$) to $11.9^{+1.4}_{-1.4}$ km \cite{Abbott_2018}. Later work constrained the properties of this event further $M_{1}=1.45^{+0.08}_{-0.06}$, $R_{1}=12.36^{+0.52}_{-0.38}$ and $M_{2}=1.28^{+0.05}_{-0.06}$, $R_{2}=12.32^{+0.66}_{-0.43}$ \cite{Fasano_2019}. Combining gravitational wave observations GW170817 and GW190425 with NICER results led to constraints on the radius of a 1.4$M_{\odot}$ of $12.33^{+0.76}_{-0.81}$ km and $12.18^{+0.56}_{-0.79}$ km \cite{Raaijmakers_2021}. 

Though GW170817 led to new constraints on the radii and tidal deformabilities of neutron stars, it alone was not sufficient to determine that the event was a binary neutron star rather than a binary black hole. The evidence that this was a binary neutron star merger came from observations of the electromagnetic counterpart. 
In future observations, we will likely not be in the fortuitous position of having a clear electromagnetic counterpart.
To date, LIGO–Virgo has detected two neutron star–black hole binaries, neither of which had an electromagnetic counterpart \cite{Abbott_2021}. Furthermore, if the mass of any of the binary components happen to lie within the mass gap, gravitational waves are the most promising avenue by which to determine whether the object is a black hole or a neutron star. 
This leads to the questions: under what conditions can a gravitational wave signal alone differentiate between a binary neutron star and a binary black hole? Can a neutron star–black hole binary be differentiated from a binary black hole by gravitational wave observations alone? This work addresses the second of these questions for current and future gravitational wave detectors. Current detectors may not be able to successfully differentiate between neutron star–black hole binaries and binary black holes, making future detectors vitally important. The importance of future detectors for studying neutron stars in binary neutron star mergers was shown in a recent paper \cite{pacilio_2021}. 

In addition to the current LIGO–Virgo detectors, we consider LIGO A+, LIGO Voyager, the Einstein Telescope, and Cosmic Explorer. The plans for LIGO A+ aim to improve the detection range of binary neutron stars at $1.4\,\mathrm{M}_{\odot}$ by a factor of 1.9 \cite{aligo}. These improvements to LIGO may occur as soon as three years from now. Further plans exist for LIGO Voyager, which will further increase detector sensitivity \cite{aligo-voyager}. Power spectral density curves for the design sensitivity of these two detectors are publicly available \cite{Ligo_ASD} and are used in our analysis. Beyond LIGO A+ and LIGO Voyager, there are plans for third-generation (3G) detectors such as the Einstein Telescope (ET) and Cosmic Explorer. We consider the Einstein Telescope and both Cosmic Explorer's first run (CE1) expected to take place in the 2030s and its second run (CE2) which is planned for the 2040s. Cosmic Explorer is expected to vastly increase the number of neutron stars detected by expanding the redshift horizon for binary neutron star detections out to 3.1 in the first run. With predicted signal-to-noise ratios going up by an order of magnitude for nearby sources, third-generation detectors will significantly improve our tidal deformability measurements \cite{Reitze_2019}. 

We use standard Bayesian model selection tools in our analysis. The evidences are calculated using the dynamic nested sampling package DYNESTY \cite{Speagle_2020} accessed through the PyCBC toolkit \cite{Biwer_2019}. In this analysis, we employ neutron star equations of state derived from chiral effective field theory, a theory that uses an effective description of nuclear matter in terms of nucleons and pions \cite{Weinberg_1990,Weinberg_1991,Machleidt_2011,Epelbaum_2009}. The chiral effective field theory framework not only leads to equations of state consistent with all symmetries of the strong interactions and known experimental constraints, but it also provides reliable uncertainty estimates. We use the same subset of the equations of state employed successfully in \cite{capano_2020} to improve constraints on neutron star radii. 

We show that, at least for the proposed gravitational wave detectors within the next decade (namely LIGO A+ and Voyager), it is very unlikely that gravitational wave observations alone will be able to distinguish neutron star–black hole binaries from binary black holes. Third-generation gravitational wave detectors will be required for this purpose.
Sec.~\ref{sec:methods} details our model selection procedure, Sec.~\ref{sec:results} presents the main results, and Sec.~\ref{sec:discussion} concludes with a discussion of the implication of these results.

\section{Methods}
\label{sec:methods}

Consider a network of gravitational wave detectors, and let $d_i(t)$ denote the gravitational wave strain time series data in the $i^{th}$ detector as a function of time $t$. The collection of all time series data in the network will be denoted $\vec{d}$. The data is the sum of detector noise $n_i(t)$ and a possible astrophysical signal $h(t)$, which depends on certain parameters which we collectively denote $\vec{\vartheta}$:
\begin{equation}
 d_i(t) = n_i(t) + h_i(t;\vec{\vartheta})\,.
\end{equation}
The central goal of a Bayesian analysis is to calculate the posterior probability distributions $p(\vec{\vartheta}| \vec{d}(t))$ of the parameters $\vec{\vartheta}$. The basis of this is Bayes' Theorem
\begin{equation} 
 p(\vec{\vartheta}| \vec{d}(t), H) = \frac{p(\vec{d}(t) | \vec{\vartheta}, H) p (\vec{\vartheta}|H) }{p(\vec{d}(t)|H)}.
\end{equation}
The fourteen parameters appearing in $\vec{\vartheta}$ are discussed below. The prior, $p(\vec{\vartheta}|H)$, represents the knowledge that we have about the parameters before considering the data. The likelihood function, $p(\vec{d}(t) | \vec{\vartheta}, H)$, is the probability of obtaining the observation $\vec{d}(t)$ given a waveform $H$ with parameters $\vec{\vartheta}$. 

In order to obtain a posterior distribution on one or a few parameters, we marginalize over the other parameters by integrating $p(\vec{d}(t) | \vec{\vartheta}, H) p (\vec{\vartheta}|H)$. Marginalizing over all parameters yields the evidence. Comparing the evidence ($p(\vec{d}(t)|H)$) of two different models ($H_{A}$ and $H_{B}$) gives the Bayes factor,
\begin{equation}
\mathcal{B} = \frac{p(\vec{d}(t)|H_{A})}{p(\vec{d}(t)|H_{B})}.
\end{equation}
This number indicates how much the data supports one model over the other. When $\mathcal{B} > 1$, $H_{A}$ is favored over $H_{B}$; the larger $\mathcal{B}$ is, the more $H_{A}$ is favored. 
In this study, the Bayes factors express how much $A$, the neutron star–black hole model, is favored over $B$, the binary black hole model. We measure evidences using the dynamic nested sampling package DYNESTY \cite{Speagle_2020}. 
To crosscheck our results, we analyze a subset of our signals using a parallel-tempered version of the emcee Markov Chain Monte Carlo sampler~\cite{Vousden:2015, emcee}. The resulting posteriors were consistent with those generated by DYNESTY.

We generate simulated gravitational waves from neutron star–black hole binary (NSBH) mergers and add these to simulated Gaussian noise colored by the target detector configuration's power spectral density (PSD). 
Gravitational waves from neutron star–black hole mergers depend on multiple variables $\vec{\vartheta}$. 
The most relevant parameters for this work are the component masses $m_{1,2}$ and the tidal deformabilities $\Lambda_{1,2}$, 
defined as
\begin{equation}
 \Lambda_{1,2} = \frac{2k_2}{3}\left( \frac{c^2R_{1,2}}{Gm_{1,2}}\right)^5.
\end{equation}
Here $R_{1,2}$ are the radii of the individual stars, and $k_2$ is the tidal Love number, which is determined from the equation of state and the mass. The leading order effect of $\Lambda_{1,2}$ on the waveform is through the combined tidal deformability parameter 
\begin{equation}
 \tilde{\Lambda} = \frac{16}{13}\frac{(12q+1)\Lambda_1 + (12+q)q^4\Lambda_2}{(1+q)^5},
\end{equation}
where we define the mass ratio $q=m_2/m_1 \geq 1$. The tidal deformability is the primary means to distinguish black holes from neutron stars using gravitational waves and infer the equation of state of neutron stars. By definition, a black hole has zero tidal deformability, while larger values of $\Lambda$ correspond to stiffer equations of state.


As the binary inspirals, merges, and then settles into a stable black hole, it emits gravitational waves. Gravitational waves have two polarizations, denoted as $h_{+,\times}$. The intrinsic parameters affect the phase evolution of the gravitational waves. Some parameters, such as the chirp mass and symmetric mass ratio, also affect the amplitude of the gravitational wave. The symmetric mass ratio $\nu$ and chirp mass $\mathcal{M}$ are defined respectively as
\begin{equation} 
 \nu = \frac{m_{1}m_{2}}{(m_{1} + m_{2})^{2}}\,,\quad \mathcal{M} = \mu^{3/5} M^{2/5}\,.
\end{equation} 

In the source frame, say one aligned with the source axis, $h_{+,\times}$ depend on the direction to the detector; equivalently, in a geocentric frame, $h_{+,\times}$ depend on the orientation of the source. 
Furthermore, the detectors do not detect $h_{+}$ and $h_{\times}$ directly, they detect the gravitational wave strain: 
\begin{equation} 
 h(t) = F_{+}(t;\alpha,\delta,\psi)h_{+} + F_{\times}(t;\alpha,\delta,\psi)h_{\times} 
\end{equation}
where $F_{+}$ and $F_{\times}$ are functions of the angles defining the location of the source. These angles are typically expressed as sky location (right ascension $\alpha$, declination $\delta$) and polarization angle $\psi$. Additionally, the amplitude depends on the inclination angle $\iota$ and the luminosity distance of the source. These extrinsic variables affect only the amplitude of the gravitational waveform. The last variable that defines the detected strain is the detection time $t_{c}$ (which determines the detector position and orientation). 

We employ the PyCBC toolkit \cite{Biwer_2019} to generate the gravitational waveforms. This requires a specification of the parameters $ \vec{\vartheta}$ and a waveform approximant. In the data analysis, colored Gaussian noise is added to the generated waveform using the detector power spectral density curve. We use a waveform approximant that combines inspiral, merger, and ringdown portions of the signal and has been calibrated to numerical relativity results (see e.g. \cite{Ajith_2008, Buonanno_1999, Damour_2010,Pannarale_2013, Lackey_2014}). The bulk of the results presented in this work use the waveform approximant SEOBNRv4\_ROM\_NRTidalv2\_NSBH \cite{Matas_2020}, which is tailored to neutron star–black hole systems. The waveform approximants IMRPhenomD\_NRTidal \cite{Khan_2016,Husa_2016,Dietrich_2017,Dietrich_2019a} and IMRPhenomNSBH \cite{Thompson:2020nei} were considered as well; however, we found SEOBNRv4\_ROM\_NRTidalv2\_NSBH to be the best choice for this analysis. We note here that neutron star–black hole systems present a considerable challenge for existing signal models and significant uncertainties remain. This is especially true at the high signal-to-noise ratios possible for third-generation detectors. For this reason, we compare the results for all three waveforms.

We set the neutron star mass to the standard $1.4 M_{\odot}$ and vary the mass of the black hole between $(5,10,15,20) M_{\odot}$ and the distance between (40,80) Mpc. 
For the neutron star, we choose two of the equations of state based on chiral effective field theory that were favored by parameter estimation in a previous work \cite{capano_2020}. The first of these equations is the maximum likelihood equation of state found therein. However, this equation of state is quite soft and leads to small tidal deformabilities ($\Lambda = 162$ for a $1.4 M_{\odot}$ neutron star). As neutron stars with large tidal deformabilities are easier to distinguish from black holes than those with small ones, we also consider a stiff equation of state. The equation of state is the stiffest equation of state in the 90th percentile credible interval of \cite{capano_2020} ($\Lambda = 369$ for a $1.4 M_{\odot}$ neutron star).

\begin{figure}[t]
 \centering
 \includegraphics[width=.85\textwidth]{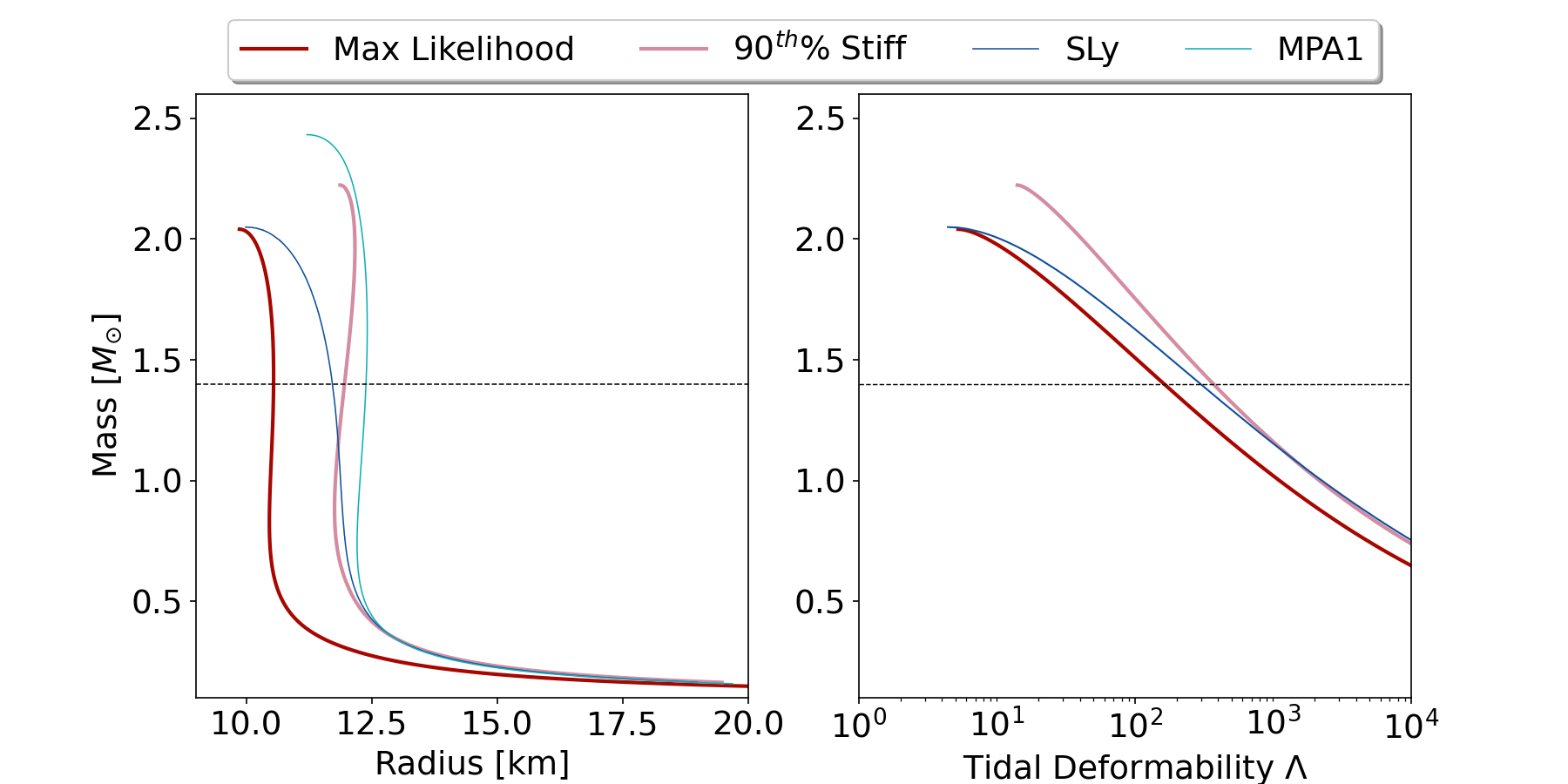}
 \caption{Mass-radius and mass-tidal deformability curves for the two equations of state used in this analysis compared to commonly used equations of state. The dashed horizontal line is at $1.4 M_{
 \odot}$. The two equations used in this paper are labelled as `stiff' and `maximum likelihood'. The stiff and maximum likelihood equations of state corresponds to $\Lambda = 369$ and $\Lambda = 162$ for a $1.4 M_{\odot}$ neutron star respectively. }
 \label{fig:mass_radius}
\end{figure}

For both the injection and the parameter estimation, the sky location is fixed to the reported sky location of GW170817 \cite{Soares_Santos_2017}:
\begin{equation}
 \alpha = 13^{\mathrm{h}}\, 09^{\mathrm{m}}\, 48.1^{\mathrm{s}}\,,\quad\delta = -23^{\circ}\, 22'\, 53.4"\,.
\end{equation}
For the injection, the polarization, inclination, and coalescence time are set to 
\begin{equation}
 \Psi = \pi\,,\quad \mathrm{tc} = 1187008882.4434\,,\quad \iota = 0.35 \,.
\end{equation}

The choice of sky location and inclination is arbitrary, and the effects of choice of sky location are discussed in Section \ref{sec:discussion}.
For the analysis with DYNESTY, we set up the parameter estimation to be as similar to the analysis for GW170817 as possible. As was done for GW170817 \cite{capano_2020,De_2018}, we fix the sky location and distance. While it is unlikely that the sky location of a detected neutron star–black hole system will be known to such accuracy, fixing the sky location in the analysis significantly reduces computation time and does not effect the resulting Bayes factors. To confirm this, we performed a series of parameter estimation runs where the sky location was a variable parameter and found the Bayes factors to be completely consistent. The variable parameters in our parameter estimation were the individual masses, spins, coalescence time, inclination, and polarization. The prior for the neutron star mass object was uniform on [1$M_{\odot}$,2$M_{\odot}$] and for the black hole it was uniform on $[m_{\mathrm{BH}}-2, m_{\mathrm{BH}}+2]$. The spin priors were both low spin $[-0.05,0.05]$, which has been used in previous analyses of GW170817 (see e.g. \cite{capano_2020,De_2018,LVC_2017}). We constrained the inclination and polarization angles to be between 0 and 2$\pi$ rad, and the coalescence time was assumed to be in the range $\mathrm{tc} \pm 0.1\,\mathrm{sec}$. 

The tidal deformability parameter estimation is what differs between our two models. To test the binary black hole hypothesis, the tidal deformability of both objects is set to 0 in the parameter estimation. We looked at two cases for the neutron star–black hole model parameter estimation. In one case, we sampled over the equation of state for the neutron star mass object. The equation of state has a uniform prior in radius at 1.4$M_{\odot}$, and there are 2,000 equations in the prior. The equation selected by the sampler was then used to calculate the tidal deformability given $m_1$ and $m_2$. In the other case, we set the equation of state as a static variable in the parameter estimation. The reason for this is that, while the nuclear equation of state is currently not well constrained, it is expected that experiments such as NICER will significantly improve our knowledge over the next decade. To take this into account, we consider the extreme case: the one in which the equation of state is known exactly and is thus fixed in the parameter estimation.

\section{Results}
\label{sec:results}

To determine if gravitational waves can distinguish between neutron star–black hole binaries and binary black holes, we look at the natural log of the Bayes factor ($\ln \mathcal{B}$) between two models. There is much debate on what constitutes evidence, strong evidence, decisive evidence, and so on. Commonly cited statistics papers such as \cite{Kass_1995} state that $\log_{10} \mathcal{B} \geq 2$ ($\ln \mathcal{B} \geq 5$) can be considered decisive evidence in favor of a model. However, this is questionable for gravitational wave model selection because of the high dimensionality, complexity and several degeneracies of the parameter space (which are not yet fully understood). Additionally, using different sampler settings and different noise realizations can lead to variations in $\ln \mathcal{B}$ of about $\pm 2$ at the $1\sigma$ level when $\ln \mathcal{B} \approx 5$, and around $\pm 4$ when $\ln \mathcal{B} \approx 10$. Taking account these uncertainties, we have decided to require a higher threshold thereby ensuring that our conclusions remain conservative regarding the capabilities of the gravitational wave detectors that we consider. We require 
\begin{equation}
\label{eq:bayesthreshold}
 \ln \mathcal{B} \geq 10
\end{equation}
for decisive evidence. 

The errors quoted in this paper are based on the standard deviation of $\ln \mathcal{B}$ across instances of the same injection parameters but with different noise realizations. Except for the specific case of $m_{\mathrm{BH}} = 5 M_{\odot}$, the errors for the current LIGO–Virgo detector network, LIGO A+, and LIGO Voyager are $<1$. For the Einstein Telescope and Cosmic Explorer 1, the errors are $<2.5$, and for Cosmic Explorer 2, the errors are $<4$. The errors in the case of $m_{\mathrm{BH}} = 5 M_{\odot}$ are larger (for details see Table \ref{Tab:errors}). The maximum error for the current LIGO–Virgo detector network is $\approx 1.0$. This increases to $2.5$ for LIGO A+, $4.7$ for LIGO Voyager, $\approx 16$ for Einstein Telescope, $\approx 19$ for Cosmic Explorer 1, and $\approx 29$ for Cosmic Explorer 2.  The relative error decreases by nearly an order of magnitude as signal-to-noise ratio increases, i.e. from ~1 for aLIGO and Virgo to ~0.1 for CE2.

\begin{figure}[ht]
\centering
\includegraphics[width=.48\textwidth]{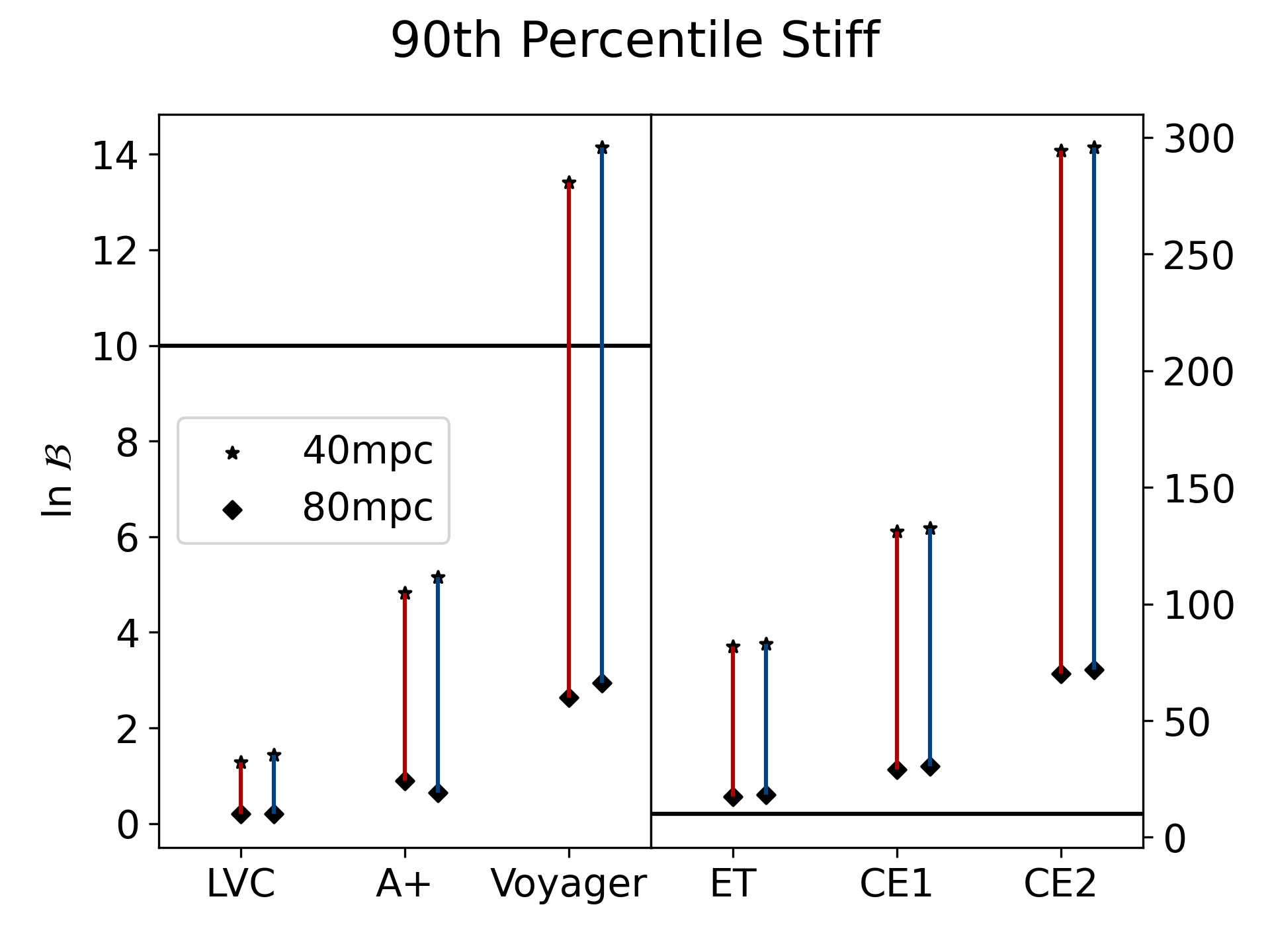}
\includegraphics[width=.48\textwidth]{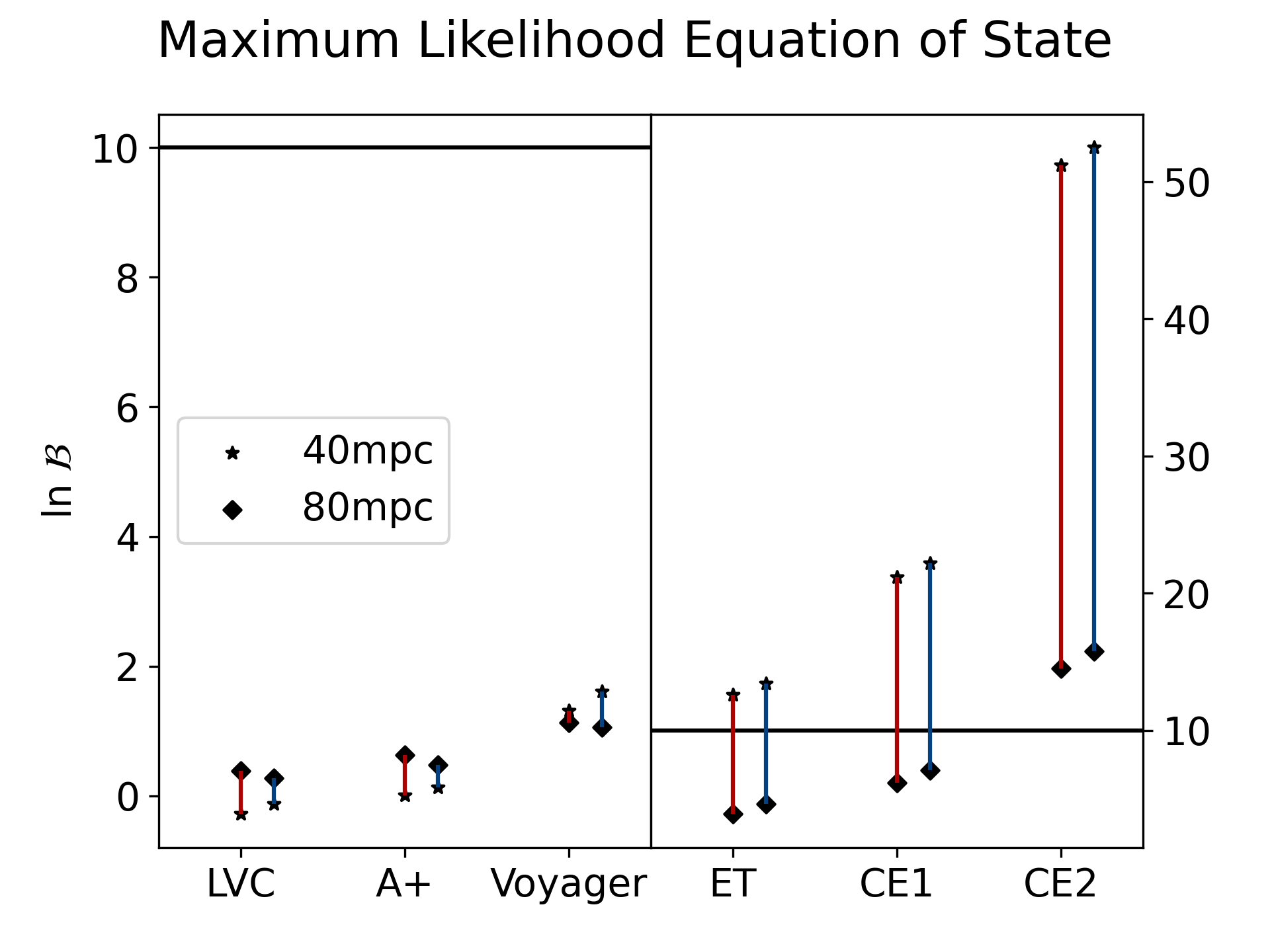}
\caption{$\ln \mathcal{B}$ for each detector with $m_{\mathrm{BH}} = 5M_{\odot}$. The vertical line spans the range of Bayes factors for a given detector. The red line on the left indicates a variable equation of state run, and the blue line on the right indicates a constant equation of state. The horizontal black line corresponds to the $\ln \mathcal{B} = 10$ cutoff. As the 3G detectors have significantly high Bayes factors, the plots are split with different $y$-axes for current and third-generation detectors.}
\label{fig:detector_Bayes_mbh5_const}
\end{figure}

\begin{figure}[ht]
\centering
\includegraphics[width=.48\textwidth]{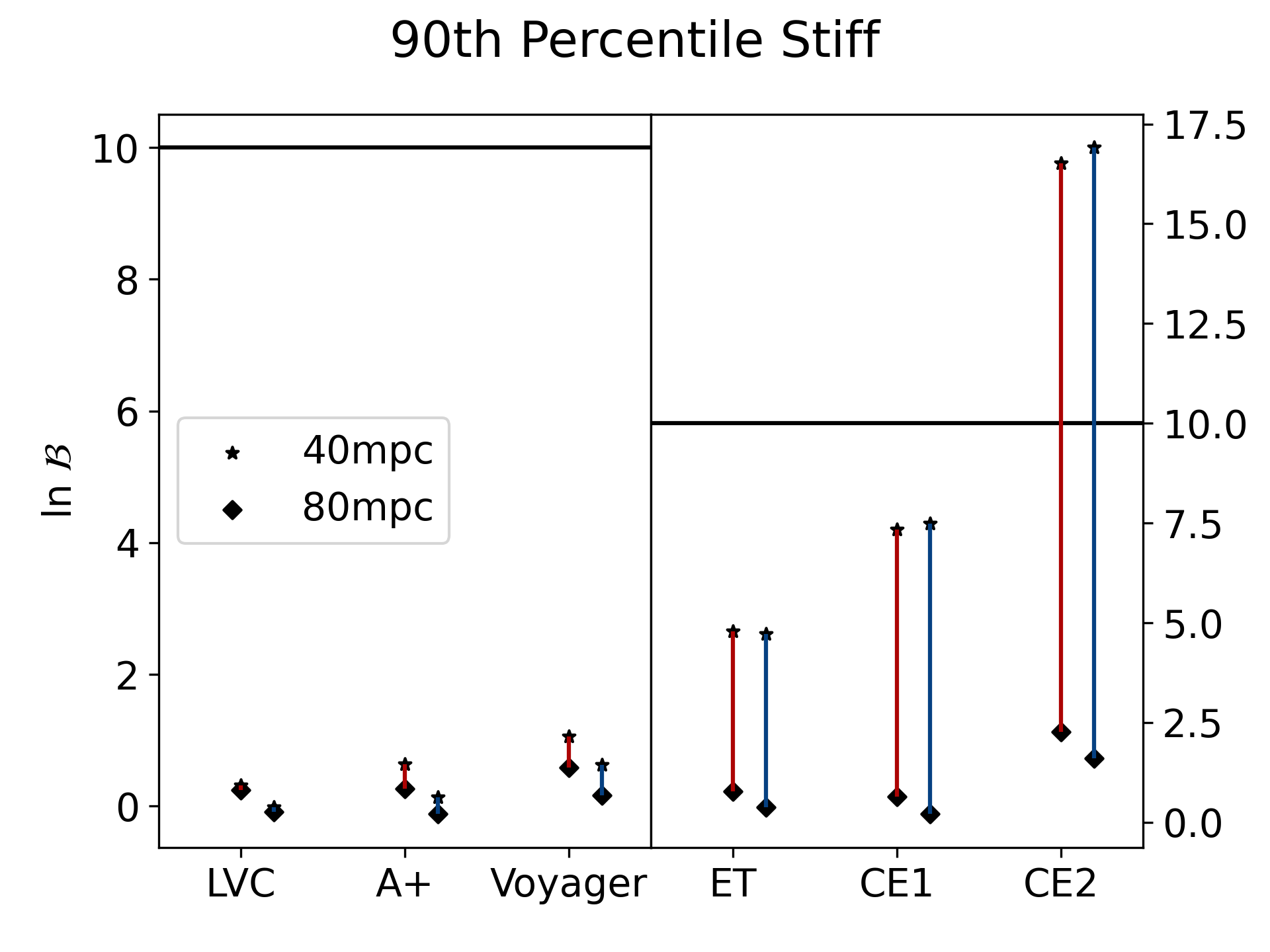}
\includegraphics[width=.48\textwidth]{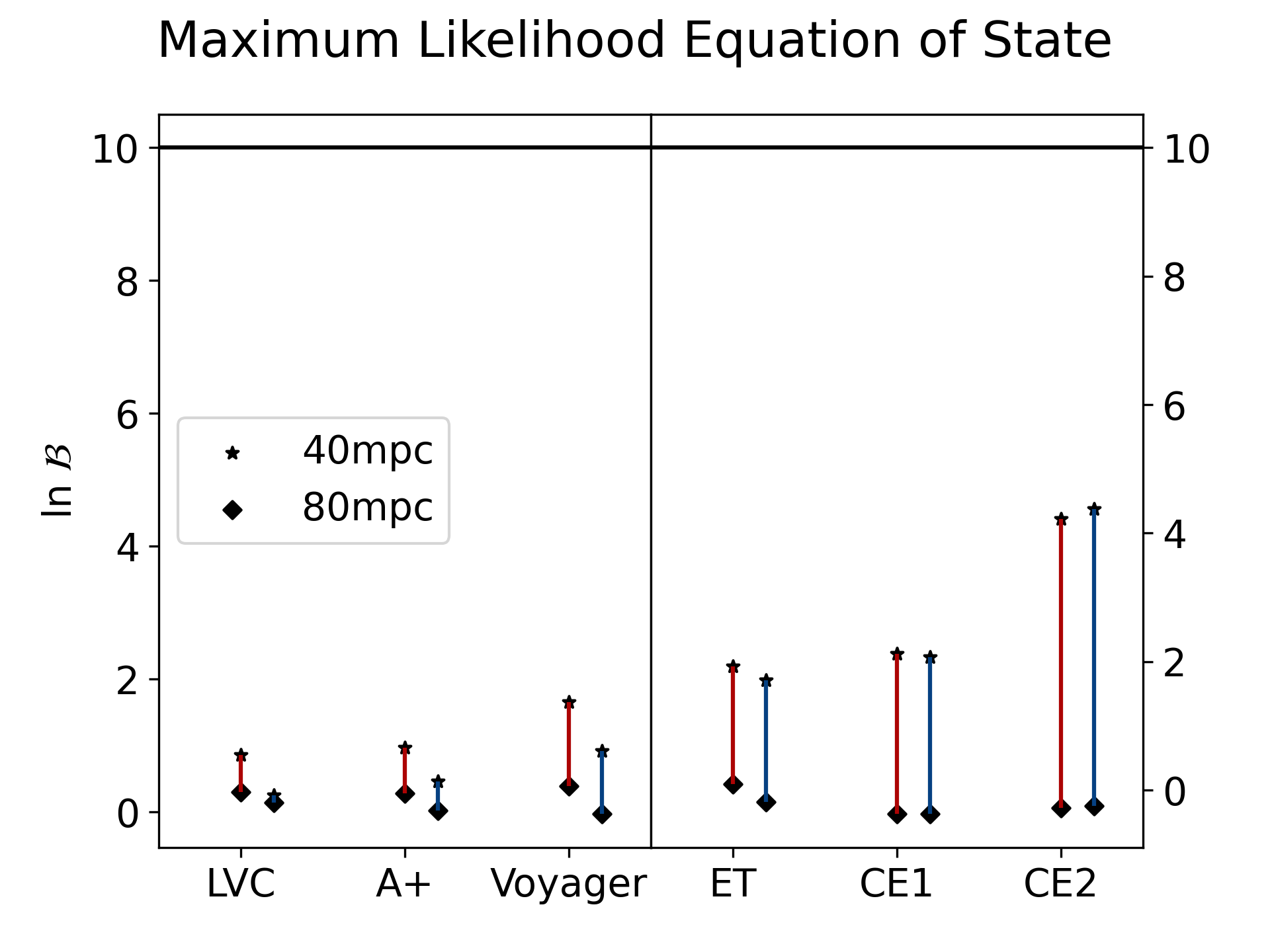}
\caption{$\ln \mathcal{B}$ for each detector with $m_{\mathrm{BH}} = 10M_{\odot}$. The vertical line spans the range of Bayes factors for a given detector. The red line on the left indicates a variable equation of state run, and the blue line on the right indicates a constant equation of state. The horizontal black line corresponds to the $\ln \mathcal{B} = 10$ cutoff.}
\label{fig:detector_Bayes_mbh10_const}
\end{figure}

\begin{figure}[!htb]
\centering
\includegraphics[width=.48\textwidth]{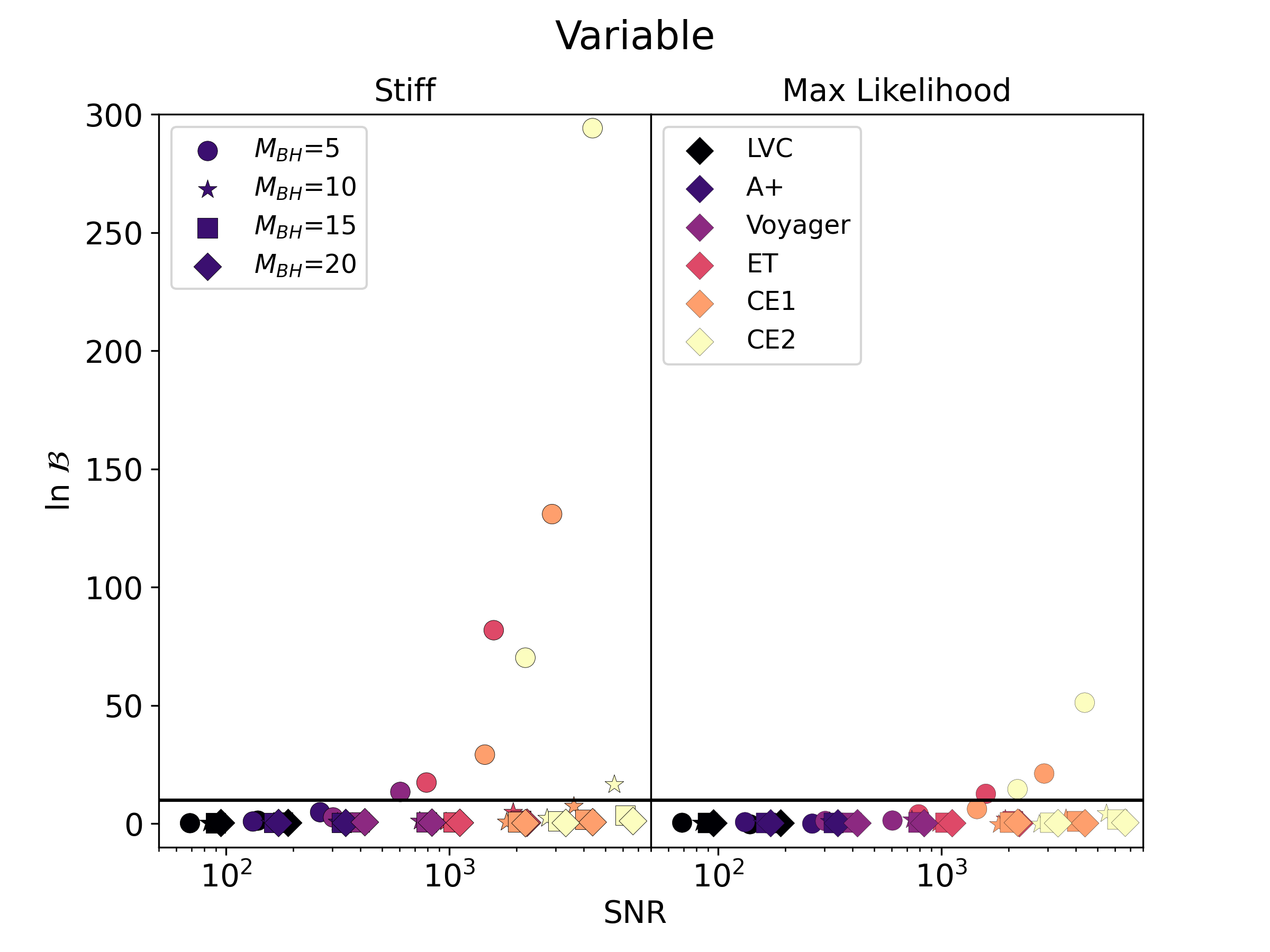}
\includegraphics[width=.48\textwidth]{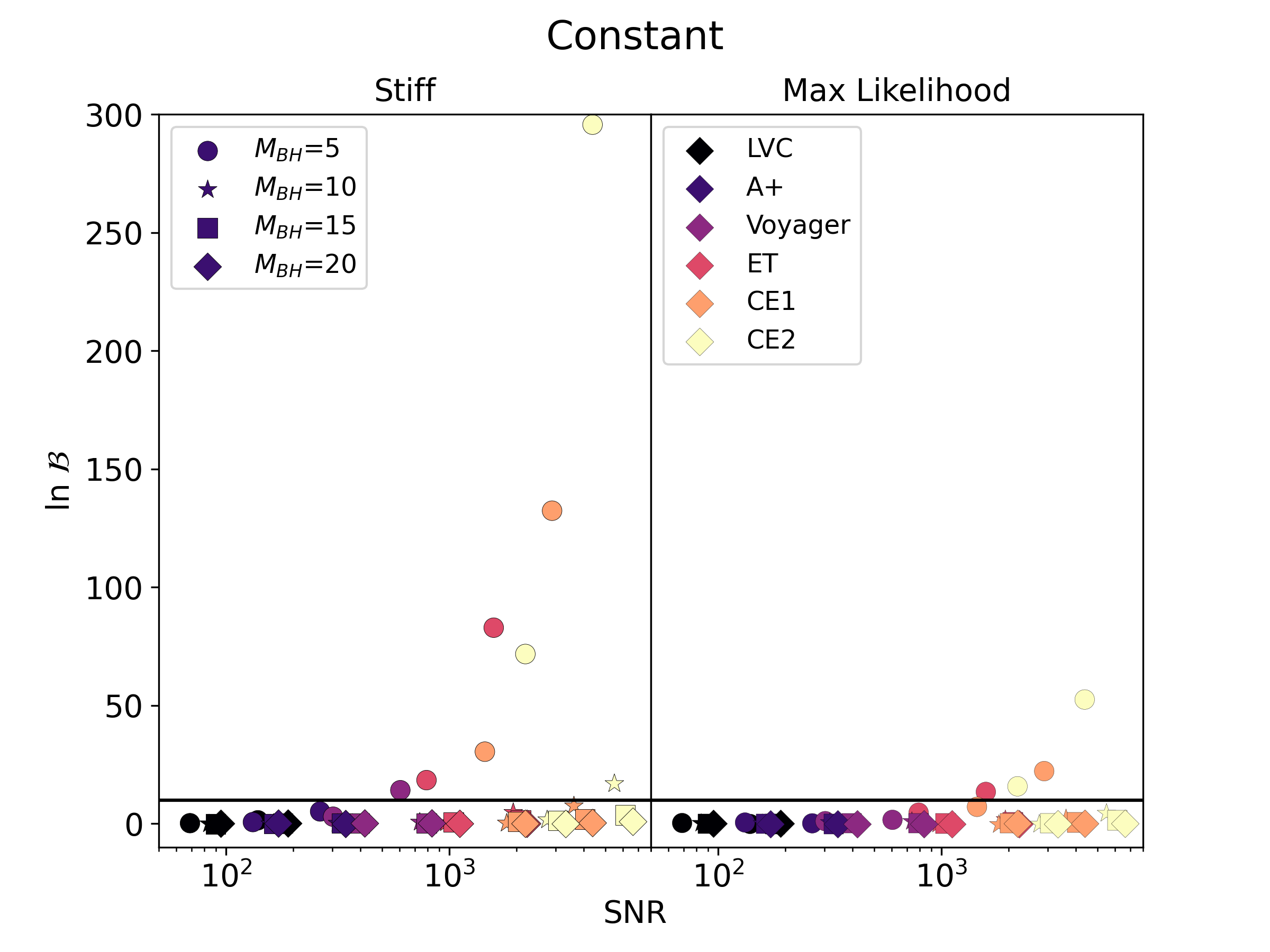}
\caption{$\ln$ Bayes Factor for all combinations of parameters and detectors as a function of the signal-to-noise ratio. The right and left plots correspond to the constant and variable equation of state cases. In each plot, the right panel shows the results for the maximum likelihood equation of state and the left panel shows the stiff equation of state. The marker color corresponds to the detector, and the marker shape indicates the mass of the black hole. The black horizontal line shows the cutoff of $\ln \mathcal{B} = 10$.}
\label{fig:snr_Bayes_phenomNSBH}
\end{figure}

As mentioned earlier, we shall present results for a $1.4M_\odot$ neutron star with a black hole companion of mass $(5,10,15,20) M_{\odot}$, and we shall take the distance to be $40\,$ or $80\,$Mpc. The neutron star equation of state shall be either of the ones shown in Fig.~\ref{fig:mass_radius}. We shall consider the following detector networks: 
\begin{itemize}
 \item The current LIGO–Virgo Network at design sensitivity in the zero-detuned high power configuration \cite{lalsim}
 \item The LIGO A+ upgrade \cite{Ligo_ASD}. 
 \item LIGO Voyager \cite{Ligo_ASD}.
 \item The Einstein Telescope \cite{Ligo_ASD}.
 \item The first observational run of the proposed $40\,$km Cosmic Explorer detector in the ``compact binary" configuration \cite{Srivastava_2020}.
 \item The second observational run of the proposed $40\,$km Cosmic Explorer detector, again in the ``compact binary" configuration \cite{Srivastava_2020}.
\end{itemize}
The results of our analysis for the various combinations of masses, distances and detector network are shown in Figs.~\ref{fig:detector_Bayes_mbh5_const}, \ref{fig:detector_Bayes_mbh10_const}, and \ref{fig:snr_Bayes_phenomNSBH}, and in Tables \ref{Tab:NSBH_LVC}-\ref{Tab:NSBH_CE2_const}. 

The Figs.~\ref{fig:detector_Bayes_mbh5_const} and \ref{fig:detector_Bayes_mbh10_const} show the Bayes factors for the $5M_\odot$ and $10M_\odot$ black hole cases respectively. As expected, the $5M_\odot$ case leads to larger Bayes factors since the tidal effects on the neutron star are more significant. Nevertheless, for both cases, the important observation for our purposes is that the Bayes factors exceed our chosen threshold of Eq.~\ref{eq:bayesthreshold} for the third-generation detectors (The Einstein Telescope and Cosmic Explorer). 
The Voyager results in Fig.~\ref{fig:detector_Bayes_mbh5_const} for the stiff equation of state surpass the threshold slightly. However, the variation is seen to be large. Furthermore, this is not the case for the maximum likelihood equation of state or for a 10 solar mass black hole companion. In a fine-tuned case, LIGO Voyager might be able to do this measurement. However, as it requires the event to be closer than any binary detected to date, to have a black hole companion that is smaller than any black hole observed by LIGO thus far, and to have a nuclear equation of state that is rather optimistically stiff, it is unlikely. This same conclusion is evident in Fig.~\ref{fig:snr_Bayes_phenomNSBH}, which shows all the combinations that we have considered: $\ln \mathcal{B}>10$ almost exclusively for the third-generation detectors. 

The precise numerical values for the Bayes factors are found in Tables \ref{Tab:NSBH_LVC}-\ref{Tab:NSBH_CE2_const}. Looking at Tables \ref{Tab:NSBH_LVC} and \ref{Tab:NSBH_LVC_const}, we see that for the current LIGO–Virgo detector network $|\ln \mathcal{B}| < 1.5$ in all cases. 
For the upgraded detector LIGO A+, the range of Bayes factors is [0.0,4.8] for the variable equation of state analysis and [-0.2,5.2] for the constant equation of state analysis. From Tables \ref{Tab:NSBH_aplus} and \ref{Tab:NSBH_aplus_const}, we see that except for the 5$M_{\odot}$ black hole companion and the $90^{th}$ percentile stiff equation of state all $|\ln\mathcal{B}| \leq 1.0$. We also see that the largest Bayes factor occurs, as expected, for the $90^{th}$ percentile stiff equation of state with a 5$M_{\odot}$ black hole at 40Mpc. 
With the LIGO Voyager, we once again see the highest Bayes factor for the 5$M_{\odot}$ black hole companion and the stiff equation of state at 40Mpc. In this case, we have 13.4 for the variable equation of state case and 14.1 for the constant equation of state case. Excluding these values, the range of $\ln \mathcal{B}$ is [0.2,2.6] for the variable equation of state and [-0.1, 2.9] for the constant equation of state case. 

The results using the third-generation detectors (the Einstein Telescope, Cosmic Explorer 1, and Cosmic Explorer 2) are more optimistic. We finally see multiple values above the threshold of 10, though in two cases, the 1$\sigma$ error falls below the cutoff. Looking at Tables \ref{Tab:NSBH_ET} and \ref{Tab:NSBH_ET_const}, we see that for the Einstein Telescope, there are now three instances for both the variable and constant equation of state cases that exceed our Bayes factor threshold. In all cases, this occurs for a black hole companion of mass 5$M_{\odot}$. For the stiff equation of state, we have $\ln \mathcal{B}$ of (81.8, 82.9) at 40Mpc and (17.4,18.4) at 80Mpc for the (variable, constant) equation of state cases. Additionally, for the maximum likelihood equation of state, at 40Mpc, we have $\ln \mathcal{B}$ of (12.6, 13.4) for the variable and constant equation of state, respectively. The results improve further when looking at Cosmic Explorer 1 and 2, particularly for the maximum likelihood equation of state. We now see multiple results above the threshold at the $1 \sigma$ level. For the stiff equation of state we have $\ln \mathcal{B}$ of (130.9,132.3) at 40Mpc and (29.2,30.4) at 80Mpc for the (variable, constant) equation of state cases. Additionally, for the maximum likelihood equation of state, at 40Mpc, we have $\ln \mathcal{B}$ of (21.2, 22.2) for the variable and constant equation of state, respectively.
In Tables \ref{Tab:NSBH_CE2} and \ref{Tab:NSBH_CE2_const}, it can be seen that the results of CE2 are very similar to those of CE1. We have $\ln \mathcal{B}> 10$ for cases with a low black hole mass, for both equations of state, out to 80Mpc for both equations of state. For CE2 we see for the first time, the possibility of distinguishing a neutron star with a $10 M_{\odot}$ black hole companion. Note, however, that this occurs only for the stiff equation of state at 40 Mpc.

\begin{figure}[ht]
\centering
\includegraphics[width=.48\textwidth]{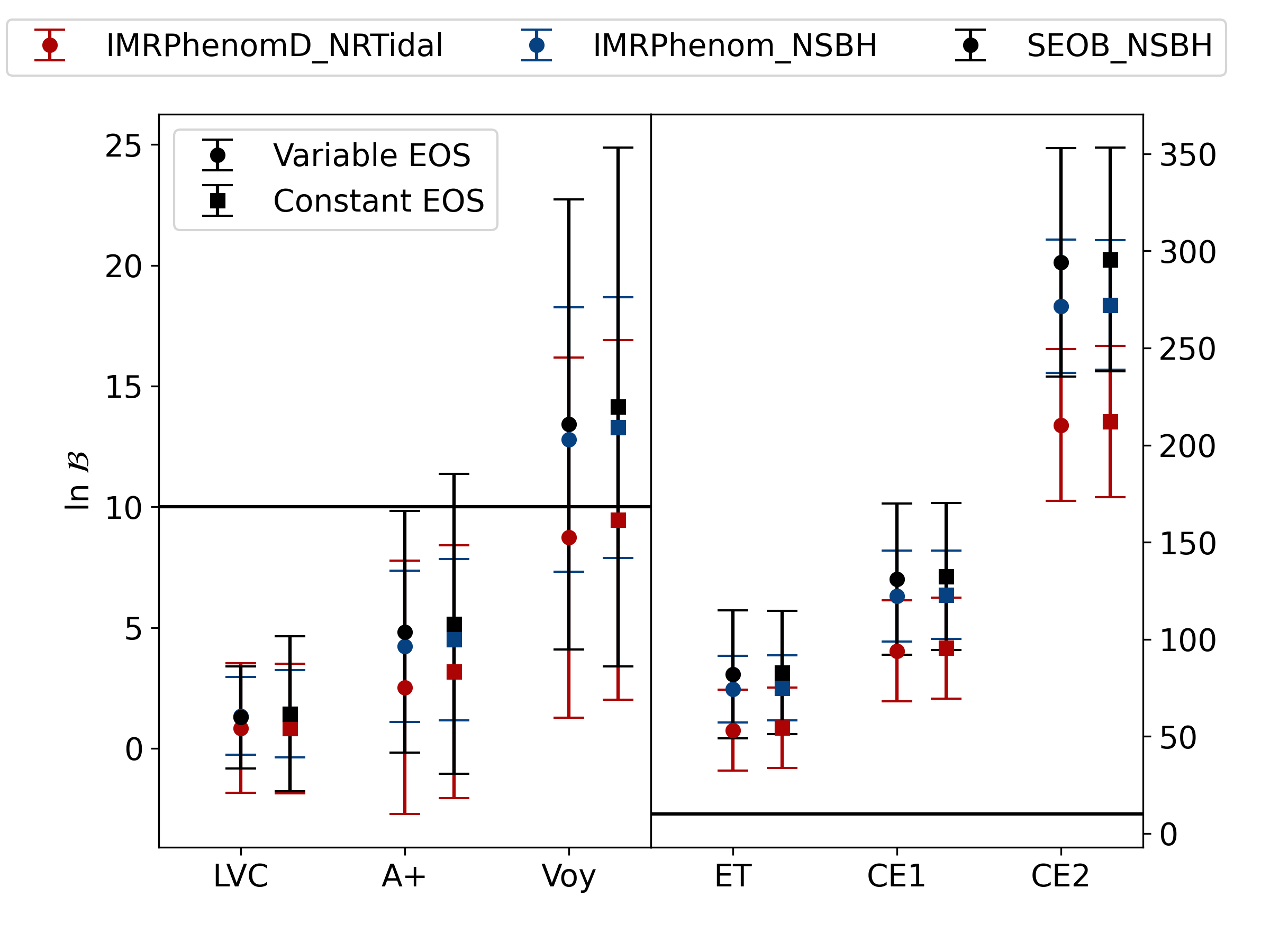}
\includegraphics[width=.48\textwidth]{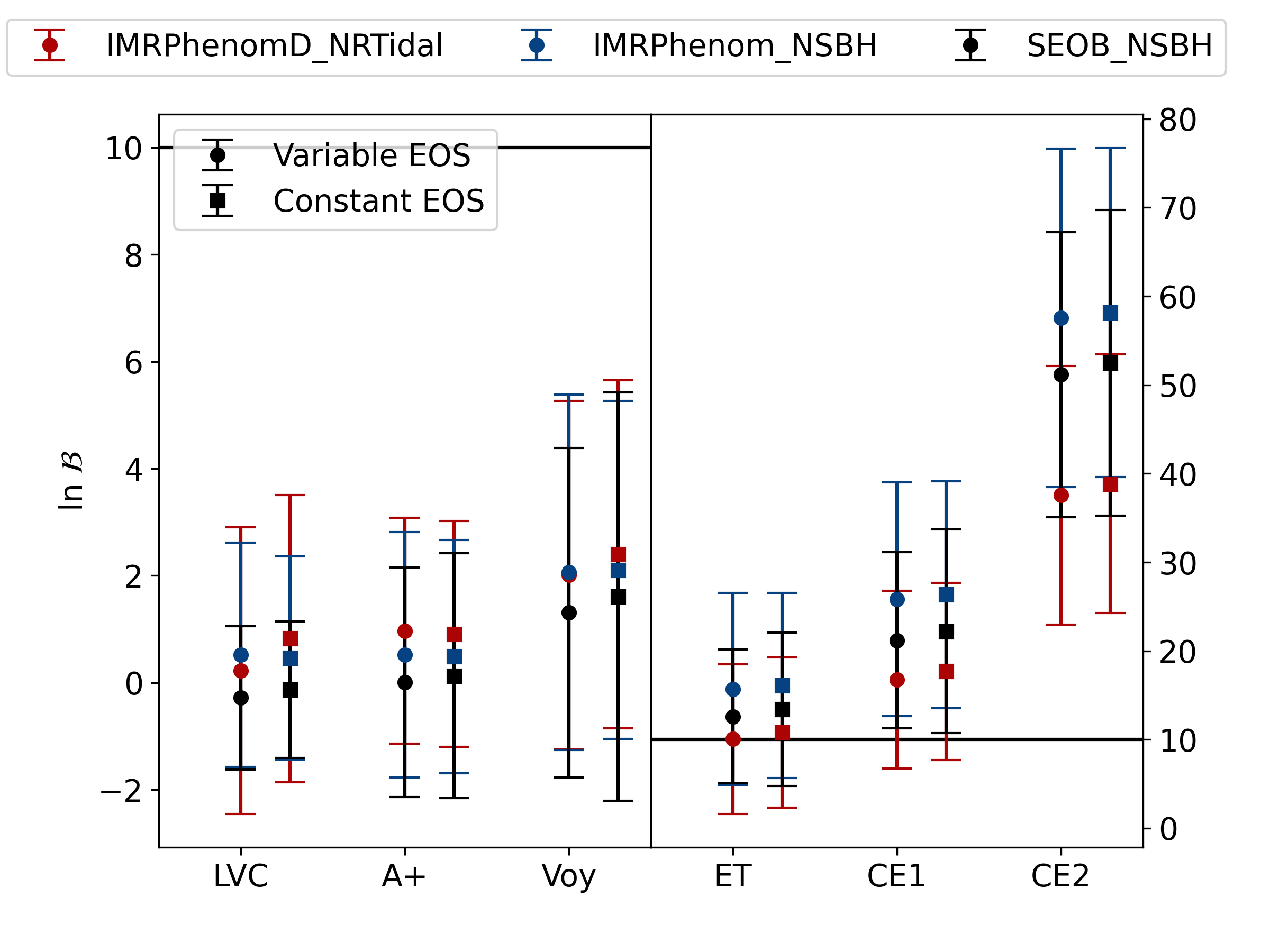}
\caption{Comparison between the IMRPhenomD\_NRTidal, IMRPhenomNSBH, and SEOBNRv4\_ROM\_NRTidalv2\_NSBH waveform approximants. For each detector, the variable equation of state analysis is on the left (circle marker) and the constant equation of state analysis is on the right (square marker). Each point corresponds to the average value of ten runs with different noise realizations, and the error bars are 2 standard deviations. The horizontal black line indicates the cutoff of $\ln \mathcal{B} = 10$.
\label{fig:compare_approximants}}
\end{figure}

Finally, we note that current waveform models for neutron star–black hole binaries have limitations in high signal-to-noise ratio and high mass ratio regimes such as the ones explored in this paper. Developing more accurate waveform models is important for analyzing real data. However, in our studies, we inject simulated signals in noise and recover them with the same signal model. Thus it is most important that the signal model capture the same qualitative features as the true signal. We considered two neutron star–black hole models, SEOBNRv4\_ROM\_NRTidalv2\_NSBH and IMRPhenomNSBH, as well as the older IMRPhenomD\_NRTidal model. Figure \ref{fig:compare_approximants} compares the results from all three waveforms for the $m_{\mathrm{BH}} = 5 M_{\odot}$ case. For LIGO–Virgo and its upgrades, all three waveforms agree at the $2 \sigma$ level. For the 3G detectors, however, IMRPhenomD\_NRTidal gives significantly lower results than the two neutron star–black hole waveforms. We see that for all three waveforms, the results qualitatively agree: The Bayes factors for Einstein Telescope and the first and second runs of Cosmic Explorer remain comfortably above the threshold.

\begin{figure}[h]
\centering
\includegraphics[width=0.5\textwidth]{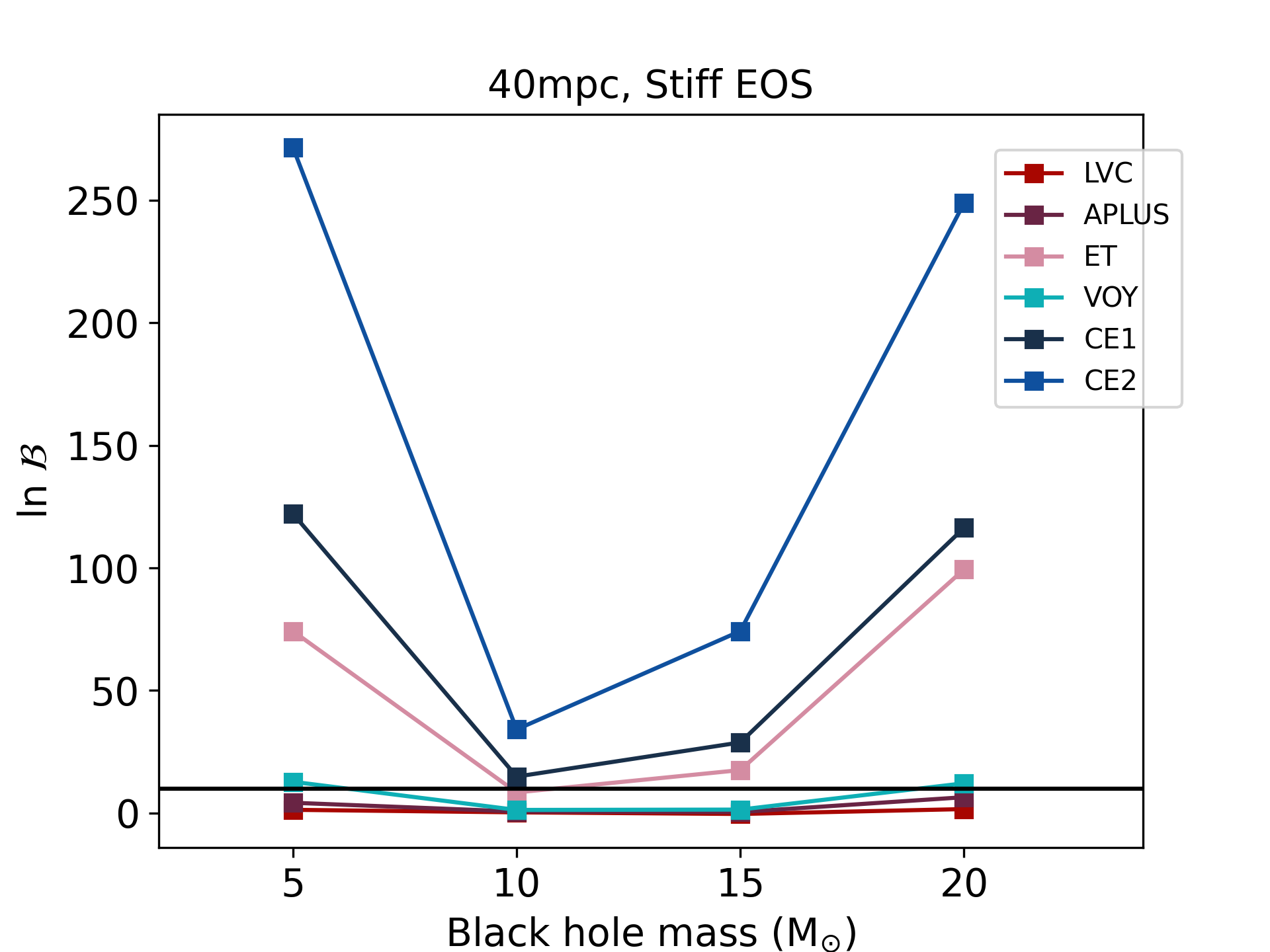}
\caption{Natural Log Bayes factor plotted as a function of black hole mass for all six detectors considered using the IMRPhenomNSBH waveform approximant. All data points are at d=40 Mpc and have a stiff equation of state ($\lambda = 369$).}
\label{fig:IMRPhenomNSBH-masslogbayes}
\end{figure}

\begin{figure}[h]
\centering
\includegraphics[width=0.6\textwidth]{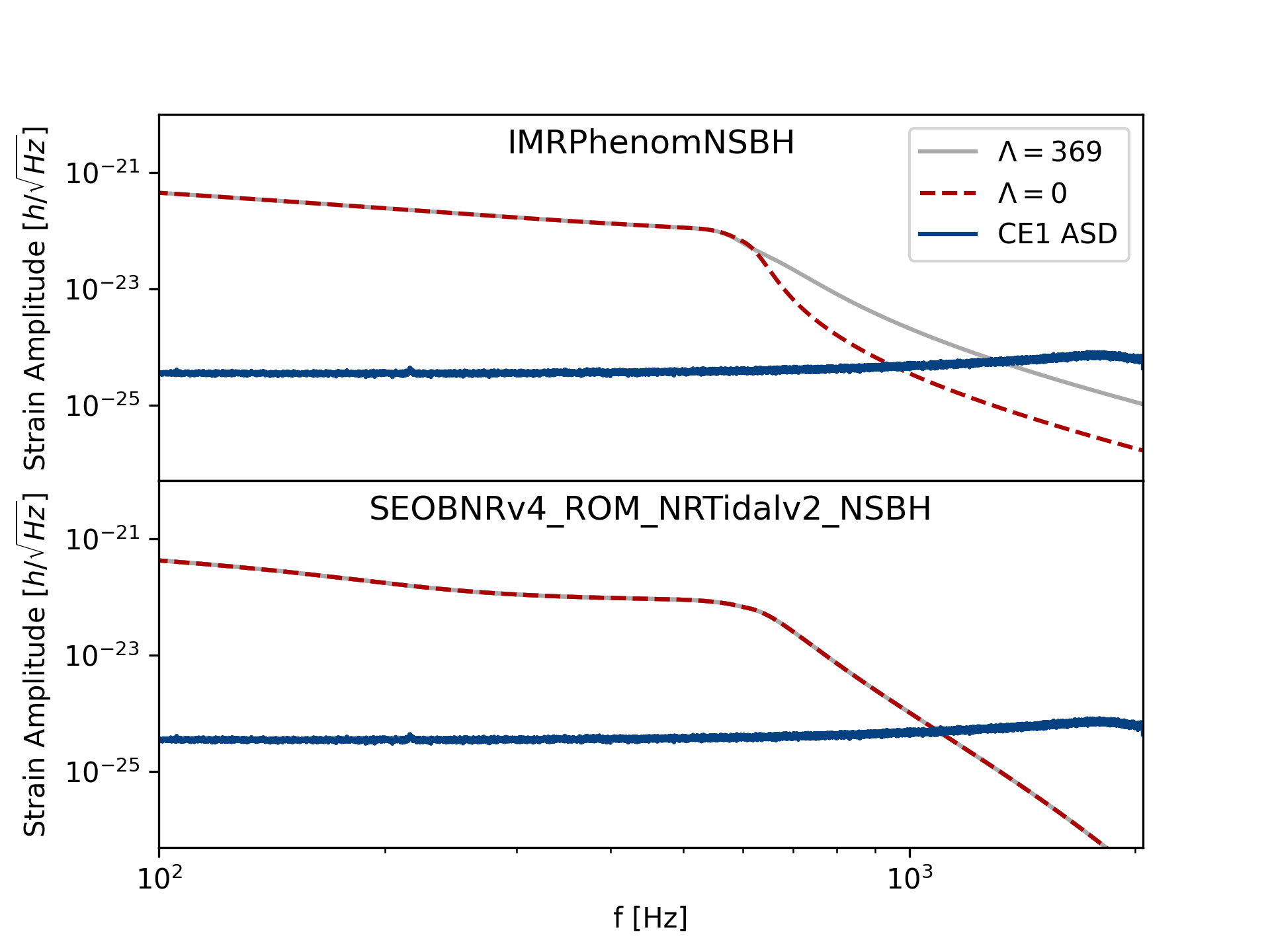}
\caption{IMRPhenomNSBH and SEOBNRv4\_ROM\_NRTidalv2\_NSBH waveform strain amplitude as a function of frequency plotted along with the Cosmic Explorer 1 amplitude spectral density. Both panels show waveforms with a black hole mass of $20 M_{\odot}$. The red line corresponds to the $\Lambda = 0$ case (the waveform of a binary black hole) \cite{Matas_2020}, and the grey line corresponds to a waveform with a stiff equation of state ( $\lambda = 369$).The top panel shows unphysical dependence of the post-merger amplitude on $\Lambda$. }
\label{fig:IMRPhenomNSBH}
\end{figure}

However, a closer look reveals that IMRPhenomNSBH is unsuitable for use with 3G detectors when $m_{\mathrm{BH}} > 10 M_{\odot}$. As the mass of the black hole companion increases, tidal effects decrease, and the gravitational waves emitted grow more similar to those of a binary black hole system. This means that the Bayes factor of the neutron star–black hole case over the binary black hole case should approach one for large black hole masses. This is indeed the behavior observed with SEOBNRv4\_ROM\_NRTidalv2\_NSBH and IMRPhenomD\_NRTidal. However, for IMRPhenomNSBH, we find that $\log \mathcal{B}$ \emph{increases} as the black hole companion mass increases from 10 to 20 $M_{\odot}$ (see Figure \ref{fig:IMRPhenomNSBH-masslogbayes}). This is clearly unphysical behavior and deserves further explanation\footnote{We thank Jonathan Thompson for discussions on this issue.}. Digging still deeper, the problem turns out to be the gravitational wave amplitude; IMRPhenomNSBH uses an older ansatz for the amplitude \cite{Santamaria:2010yb} (which was not originally intended for neutron star - black hole systems). Figure \ref{fig:IMRPhenomNSBH} compares the gravitational wave amplitude as a function of frequency for two different values of $\Lambda$, for the $m_{\mathrm{BH}} = 20 M_{\odot}$ case, for both approximants (along with the amplitude spectral density for the CE1 detector). We clearly see that while the SEOBNRv4\_ROM\_NRTidalv2\_NSBH model shows no dependence of the amplitude on $\Lambda$, which is what we expect for these high mass configurations, the IMRPhenomNSBH model shows a large dependence on $\Lambda$ for the post-merger signal which is clearly unphysical. This incorrect behavior explains the effect shown in Figure \ref{fig:IMRPhenomNSBH-masslogbayes}. Due to this non-physical behavior, we choose to use SEOBNRv4\_ROM\_NRTidalv2\_NSBH for our analysis.


\section{Discussion} 
\label{sec:discussion}

The results demonstrate that the current LIGO and Virgo detectors are not sufficient to differentiate between neutron star–black hole and binary black hole systems. In fact, the success of A+ and Voyager for this purpose is dubious. There were no cases for either LIGO–Virgo or LIGO A+ where the $\ln \mathcal{B}$ exceeded our threshold, and only a fine-tuned case for LIGO Voyager. However, it is important to note that the cases with the highest $\ln \mathcal{B}$ always occur with the stiff equation of state, the 5$M_{\odot}$ black hole companion, and at 40Mpc. This is not surprising. The stiff equation of state was selected specifically for this property, and the signal-to-noise ratio at 40Mpc is higher than at 80Mpc. 

It can be seen that the ability to differentiate between a neutron star–black hole system and a binary black hole system does not directly correspond to the signal-to-noise ratio. The highest $\ln \mathcal{B}$ occurs for $m_{\mathrm{BH}} = 5 M_{\odot}$, even though systems with $m_{\mathrm{BH}} = 10,15,20 M_{\odot}$ have higher signal-to-noise ratios. The tidal effects decrease as mass increases, and this effect is clearly of greater importance than the increase in signal strength. Detection of a neutron star–black hole system with a low mass ratio will almost certainly be required to give evidence of neutron star matter in the gravitational wave signal. Additionally, the nuclear equation of state itself is an important factor in how soon we will and how likely we are to distinguish a neutron star–black hole system from a binary black hole system. Finally, in this analysis, we have made a particular choice of sky location and inclination angle of the source. This orientation, corresponding to GW170817, is a favorable one. As expected, repeating the simulations with randomly chosen sky-positions generally leads to smaller Bayes factors. However, even in this case, the Bayes factors for the Einstein Telescope, Cosmic Explorer 1 and Cosmic Explorer 2 remain comfortably above the threshold, while for Voyager, the results get closer to the threshold. Our basic conclusions therefore remain unchanged. 

When looking at LIGO Voyager, we saw that in one case the results were close to our cutoff. Keep in mind, however, that this analysis was done with design sensitivity curves and this result occurs only for the fine-tuned case of a very close binary with a very small black hole that has a rather stiff equation of state. Despite LIGO–Virgo's recent detection of an object in the mass gap, 5$M_{\odot}$ is still on the low end of what we expect for black hole masses. Looking at Figure \ref{fig:detector_Bayes_mbh10_const}, it's evident that, when the companion mass increases to even 10$M_{\odot}$, the Bayes factor drops rapidly regardless of distance or equation of state for all detectors. If the equation of state is as soft as the analysis of GW170817 suggests, then LIGO Voyager will certainly be unable to distinguish neutron star–black hole systems from binary black holes regardless of how close or loud the signal is.

3G detectors will likely be required to obtain decisive evidence of neutron star–black hole system from gravitational wave data. We see here that the proposed designs for the Einstein Telescope and Cosmic Explorer may very well allow for these detections. Looking at Figure \ref{fig:detector_Bayes_mbh5_const} and Table \ref{Tab:NSBH_CE1}, we see that regardless of the nuclear equation of state, there are systems which have $\ln \mathcal{B} > 10$. Thus, if current analyses of GW170817 are accurate, we will be waiting until the Einstein Telescope or Cosmic Explorer for gravitational wave evidence of neutron star–black hole systems. Additionally, 3G detectors seem to be able to do this measurement at distances out 80 Mpc (stiff equation of state), which greatly expands the number of candidate systems.
Even with very sensitive future detectors, the ability to distinguish neutron star–black hole systems from binary black holes is very dependent on the mass of the black hole in the binary. 

\bibliography{SensitivitySearch}

 \bibliographystyle{aasjournal.bst}

\section*{Acknowledgements}

We thank Jonathon Thompson, Sumit Kumar, Sanjay Reddy, Ingo Tews, and Duncan Brown for their
valuable discussions. Our computations used the ATLAS computing cluster 
at AEI Hannover \cite{Atlas} funded by the Max Planck Society and the
State of Niedersachsen, Germany.

This research has made use of data, software, and/or web tools obtained from the LIGO Open Science Center (\url{https://losc.ligo.org}), a service of LIGO Laboratory, the LIGO Scientific Collaboration and the Virgo Collaboration. LIGO is funded by the U.S. National Science Foundation. Virgo is funded by the French Centre National de Recherche Scientifique (CNRS), the Italian Istituto Nazionale della Fisica Nucleare (INFN), and the Dutch Nikhef, with contributions by Polish and Hungarian institutes.

\appendix

\section{Data Tables}

\begin{table}[ht] 
\centering 
\caption{Log Bayes Factor and standard deviation for selected cases.} 
\begin{tabular}{|c|c|c|c|c|c|c|c|c|c|} 
\hline
M$_{\mathrm{BH}}$ [$M_{\odot}$] & d[Mpc] & EOS & LVC & A+ & Voyager & ET & CE1 & CE2 \\ \hline \hline
 5 & 40 & stiff & $1.3 \pm 1.0$ & $4.8 \pm 2.5$ & $13.4 \pm 4.7$ & $81.8 \pm 16.5$ & $130.9 \pm 19.0$ & $294.4 \pm 29.4$ \\ \hline 
5 & 40 & soft & $-0.3 \pm 0.7$ & $0.0 \pm 1.1$ & $1.3 \pm 1.5$ & $12.6 \pm 3.8$ & $21.2 \pm 5.0$ & $51.2 \pm 8.0$ \\ \hline 
5 & 80 & stiff & $0.2 \pm 0.5$ & $0.9 \pm 0.7$ & $2.6 \pm 1.2$ & $17.4 \pm 4.1$ & $29.2 \pm 5.6$ & $70.2 \pm 8.8$\\ \hline
5 & 80 & soft & $0.4 \pm 0.6$ & $0.6 \pm 0.9$ & $1.1 \pm 1.7$ & $3.9 \pm 1.8$ & $6.2 \pm 2.2$ & $14.6 \pm 3.3$ \\ \hline
10 & 40 & stiff & $0.3 \pm 0.2$ & $0.6 \pm 0.4 $ & $1.1 \pm 0.8$ & $4.8 \pm 2.3$ & $7.3 \pm 2.3$ & $16.5 \pm 3.8$ \\ \hline 
\end{tabular}
\label{Tab:errors}
\end{table}

\begin{table}[ht]
\caption{neutron star–black hole with LIGO–Virgo}
\begin{tabular}{|c|c|c|c||c|c|c|c|}
\hline\multicolumn{4}{|c||}{ $90^{th} \%$ Stiff } & \multicolumn{4}{|c|}{ Maximum Likelihood } \\ \hline
Mass$_{\mathrm{BH}}$ [$M_{\odot}$] & Distance [Mpc] & SNR & $\ln \mathcal{B}$ & Mass$_{\mathrm{BH}}$ [$M_{\odot}$] & Distance [Mpc] & SNR & $\ln \mathcal{B}$ \\ \hline
5 & 40 & 139 & 1.3 & 5 & 40 & 139 & -0.3 \\ \hline 
10 & 40 & 168 & 0.3 & 10 & 40 & 168 & 0.9 \\ \hline 
15 & 40 & 180 & 0.6 & 15 & 40 & 180 & 0.1 \\ \hline 
20 & 40 & 190 & 0.3 & 20 & 40 & 190 & 0.3 \\ \hline 
5 & 80 & 69 & 0.2 & 5 & 80 & 69 & 0.4 \\ \hline 
10 & 80 & 84 & 0.2 & 10 & 80 & 84 & 0.3 \\ \hline 
15 & 80 & 90 & 0.3 & 15 & 80 & 90 & 0.3 \\ \hline 
20 & 80 & 95 & 0.3 & 20 & 80 & 95 & 0.2 \\ \hline 
\end{tabular}
\label{Tab:NSBH_LVC}
\end{table}

\begin{table}[ht]
\caption{neutron star–black hole with LIGO–Virgo and equation of state held constant }
\begin{tabular}{|c|c|c|c||c|c|c|c|}
\hline\multicolumn{4}{|c||}{ $90^{th} \%$ Stiff } & \multicolumn{4}{|c|}{ Maximum Likelihood } \\ \hline
Mass$_{\mathrm{BH}}$ [$M_{\odot}$] & Distance [Mpc] & SNR & $\ln\mathcal{B}$ & Mass$_{\mathrm{BH}}$ [$M_{\odot}$] & Distance [Mpc] & SNR & $\ln\mathcal{B}$ \\ \hline
5 & 40 & 139 & 1.4 & 5 & 40 & 139 & -0.1 \\ \hline 
10 & 40 & 168 & 0.0 & 10 & 40 & 168 & 0.2 \\ \hline 
15 & 40 & 180 & 0.2 & 15 & 40 & 180 & -0.2 \\ \hline 
20 & 40 & 190 & 0.0 & 20 & 40 & 190 & 0.1 \\ \hline 
5 & 80 & 69 & 0.2 & 5 & 80 & 69 & 0.3 \\ \hline 
10 & 80 & 84 & -0.1 & 10 & 80 & 84 & 0.1 \\ \hline 
15 & 80 & 90 & -0.2 & 15 & 80 & 90 & 0.0 \\ \hline 
20 & 80 & 95 & 0.0 & 20 & 80 & 95 & 0.0 \\ \hline 

\end{tabular}
\label{Tab:NSBH_LVC_const}
\end{table}

\begin{table}[ht]
\caption{neutron star–black hole with LIGO A+}
\begin{tabular}{|c|c|c|c||c|c|c|c|}
\hline
\multicolumn{4}{|c||}{ $90^{th} \%$ Stiff } & \multicolumn{4}{|c|}{ Maximum Likelihood } \\ \hline
Mass$_{\mathrm{BH}}$ [$M_{\odot}$] & Distance [Mpc] & SNR & $\ln\mathcal{B}$ & Mass$_{\mathrm{BH}}$ [$M_{\odot}$] & Distance [Mpc] & SNR & $\ln\mathcal{B}$ \\ \hline
5 & 40 & 264 & 4.8 & 5 & 40 & 264 & 0.0 \\ \hline 
10 & 40 & 316 & 0.6 & 10 & 40 & 316 & 1.0 \\ \hline 
15 & 40 & 329 & 0.5 & 15 & 40 & 329 & 0.3 \\ \hline 
20 & 40 & 344 & 0.3 & 20 & 40 & 344 & 0.4 \\ \hline 
5 & 80 & 132 & 0.9 & 5 & 80 & 132 & 0.6 \\ \hline 
10 & 80 & 158 & 0.3 & 10 & 80 & 158 & 0.3 \\ \hline 
15 & 80 & 164 & 0.4 & 15 & 80 & 164 & 0.3 \\ \hline 
20 & 80 & 172 & 0.3 & 20 & 80 & 172 & 0.3 \\ \hline 
\end{tabular}
\label{Tab:NSBH_aplus}
\end{table}

\begin{table}[ht]
\caption{neutron star–black hole with LIGO A+ with equation of state held constant}
\begin{tabular}{|c|c|c|c||c|c|c|c|}
\hline
\multicolumn{4}{|c||}{ $90^{th} \%$ Stiff } & \multicolumn{4}{|c|}{ Maximum Likelihood } \\ \hline
Mass$_{\mathrm{BH}}$ [$M_{\odot}$] & Distance [Mpc] & SNR & $\ln\mathcal{B}$ & Mass$_{\mathrm{BH}}$ [$M_{\odot}$] & Distance [Mpc] & SNR & $\ln\mathcal{B}$ \\ \hline
5 & 40 & 264 & 5.2 & 5 & 40 & 264 & 0.1 \\ \hline 
10 & 40 & 316 & 0.1 & 10 & 40 & 316 & 0.5 \\ \hline 
15 & 40 & 329 & 0.1 & 15 & 40 & 329 & -0.2 \\ \hline 
20 & 40 & 344 & -0.2 & 20 & 40 & 344 & -0.2 \\ \hline 
5 & 80 & 132 & 0.6 & 5 & 80 & 132 & 0.5 \\ \hline 
10 & 80 & 158 & -0.1 & 10 & 80 & 158 & 0.0 \\ \hline 
15 & 80 & 164 & -0.1 & 15 & 80 & 164 & -0.1 \\ \hline 
20 & 80 & 172 & 0.0 & 20 & 80 & 172 & -0.1 \\ \hline 
\end{tabular} 

\label{Tab:NSBH_aplus_const}
\end{table}

\begin{table}[ht]
\caption{neutron star–black hole with LIGO Voyager}
\begin{tabular}{|c|c|c|c||c|c|c|c|}
\hline
\multicolumn{4}{|c||}{ $90^{th} \%$ Stiff } & \multicolumn{4}{|c|}{ Maximum Likelihood } \\ \hline
Mass$_{\mathrm{BH}}$ [$M_{\odot}$] & Distance [Mpc] & SNR & $\ln\mathcal{B}$ & Mass$_{\mathrm{BH}}$ [$M_{\odot}$] & Distance [Mpc] & SNR & $\ln\mathcal{B}$ \\ \hline
5 & 40 & 604 & 13.4 & 5 & 40 & 604 & 1.3 \\ \hline 
10 & 40 & 738 & 1.1 & 10 & 40 & 738 & 1.6 \\ \hline 
15 & 40 & 791 & 0.6 & 15 & 40 & 791 & 0.5 \\ \hline 
20 & 40 & 837 & 0.4 & 20 & 40 & 837 & 0.2 \\ \hline 
5 & 80 & 302 & 2.6 & 5 & 80 & 302 & 1.1 \\ \hline 
10 & 80 & 369 & 0.6 & 10 & 80 & 369 & 0.4 \\ \hline 
15 & 80 & 396 & 0.6 & 15 & 80 & 396 & 0.4 \\ \hline 
20 & 80 & 419 & 0.6 & 20 & 80 & 419 & 0.2 \\ \hline 
\end{tabular}
\label{Tab:NSBH_voyager}
\end{table}

\begin{table}[ht]
\caption{neutron star–black hole with LIGO Voyager and equation of state held constant}
\begin{tabular}{|c|c|c|c||c|c|c|c|}
\hline
\multicolumn{4}{|c||}{ $90^{th} \%$ Stiff } & \multicolumn{4}{|c|}{ Maximum Likelihood } \\ \hline
Mass$_{\mathrm{BH}}$ [$M_{\odot}$] & Distance [Mpc] & SNR & $\ln\mathcal{B}$ & Mass$_{\mathrm{BH}}$ [$M_{\odot}$] & Distance [Mpc] & SNR & $\ln\mathcal{B}$ \\ \hline
5 & 40 & 604 & 14.1 & 5 & 40 & 604 & 1.6 \\ \hline 
10 & 40 & 738 & 0.6 & 10 & 40 & 738 & 0.9 \\ \hline 
15 & 40 & 791 & 0.2 & 15 & 40 & 791 & 0.1 \\ \hline 
20 & 40 & 837 & 0.1 & 20 & 40 & 837 & -0.1 \\ \hline 
5 & 80 & 302 & 2.9 & 5 & 80 & 302 & 1.1 \\ \hline 
10 & 80 & 369 & 0.2 & 10 & 80 & 369 & -0.0 \\ \hline 
15 & 80 & 396 & 0.2 & 15 & 80 & 396 & 0.2 \\ \hline 
20 & 80 & 419 & 0.1 & 20 & 80 & 419 & -0.1 \\ \hline 

\end{tabular} 
\label{Tab:NSBH_voyager_const}
\end{table}

\begin{table}[ht] 
\caption{neutron star–black hole with Einstein Telescope} 
\begin{tabular}{|c|c|c|c||c|c|c|c|} 
\hline\multicolumn{4}{|c||}{ $90^{th} \%$ Stiff } & \multicolumn{4}{|c|}{ Maximum Likelihood } \\ \hline 
Mass$_{\mathrm{BH}}$ [$M_{\odot}$] & Distance [Mpc] & SNR & $\ln\mathcal{B}$ & Mass$_{\mathrm{BH}}$ [$M_{\odot}$] & Distance [Mpc] & SNR & $\ln\mathcal{B}$ \\ \hline 5 & 40 & 1582 & 81.8 & 5 & 40 & 1582 & 12.6 \\ \hline 
10 & 40 & 1935 & 4.8 & 10 & 40 & 1935 & 1.9 \\ \hline 
15 & 40 & 2091 & 1.3 & 15 & 40 & 2091 & 1.0 \\ \hline 
20 & 40 & 2233 & 0.4 & 20 & 40 & 2233 & 0.3 \\ \hline 
5 & 80 & 791 & 17.4 & 5 & 80 & 791 & 3.9 \\ \hline 
10 & 80 & 968 & 0.8 & 10 & 80 & 968 & 0.1 \\ \hline 
15 & 80 & 1045 & 0.6 & 15 & 80 & 1045 & 0.4 \\ \hline 
20 & 80 & 1116 & 0.4 & 20 & 80 & 1116 & 0.2 \\ \hline 
\end{tabular} 
\label{Tab:NSBH_ET}
\end{table} 

\begin{table}[ht] 
\caption{neutron star–black hole with Einstein Telescope and equation of state held constant} 
\begin{tabular}{|c|c|c|c||c|c|c|c|} 
\hline\multicolumn{4}{|c||}{ $90^{th} \%$ Stiff } & \multicolumn{4}{|c|}{ Maximum Likelihood } \\ \hline 
Mass$_{\mathrm{BH}}$ [$M_{\odot}$] & Distance [Mpc] & SNR & $\ln\mathcal{B}$ & Mass$_{\mathrm{BH}}$ [$M_{\odot}$] & Distance [Mpc] & SNR & $\ln\mathcal{B}$ \\ \hline 5 & 40 & 1582 & 82.9 & 5 & 40 & 1582 & 13.4 \\ \hline 
10 & 40 & 1935 & 4.7 & 10 & 40 & 1935 & 1.7 \\ \hline 
15 & 40 & 2091 & 1.6 & 15 & 40 & 2091 & 0.6 \\ \hline 
20 & 40 & 2233 & -0.0 & 20 & 40 & 2233 & -0.1 \\ \hline 
5 & 80 & 791 & 18.4 & 5 & 80 & 791 & 4.6 \\ \hline 
10 & 80 & 968 & 0.4 & 10 & 80 & 968 & -0.2 \\ \hline 
15 & 80 & 1045 & 0.6 & 15 & 80 & 1045 & 0.1 \\ \hline 
20 & 80 & 1116 & -0.1 & 20 & 80 & 1116 & -0.1 \\ \hline 
\end{tabular} 
\label{Tab:NSBH_ET_const}
\end{table} 

\begin{table}[ht] 
\caption{neutron star–black hole with Cosmic Explorer 1 40km} 
\begin{tabular}{|c|c|c|c||c|c|c|c|} 
\hline\multicolumn{4}{|c||}{ $90^{th} \%$ Stiff } & \multicolumn{4}{|c|}{ Maximum Likelihood } \\ \hline 
Mass$_{\mathrm{BH}}$ [$M_{\odot}$] & Distance [Mpc] & SNR & $\ln\mathcal{B}$ & Mass$_{\mathrm{BH}}$ [$M_{\odot}$] & Distance [Mpc] & SNR & $\ln\mathcal{B}$ \\ \hline 5 & 40 & 2888 & 130.9 & 5 & 40 & 2888 & 21.2 \\ \hline 
10 & 40 & 3617 & 7.3 & 10 & 40 & 3617 & 2.1 \\ \hline 
15 & 40 & 4055 & 1.8 & 15 & 40 & 4055 & 1.0 \\ \hline 
20 & 40 & 4390 & 0.6 & 20 & 40 & 4390 & 0.2 \\ \hline 
5 & 80 & 1444 & 29.2 & 5 & 80 & 1444 & 6.2 \\ \hline 
10 & 80 & 1809 & 0.6 & 10 & 80 & 1809 & -0.4 \\ \hline 
15 & 80 & 2028 & 0.8 & 15 & 80 & 2028 & 0.5 \\ \hline 
20 & 80 & 2195 & 0.2 & 20 & 80 & 2195 & 0.5 \\ \hline 
\end{tabular} 
\label{Tab:NSBH_CE1}
\end{table} 

\begin{table}[ht] 
\caption{neutron star–black hole with Cosmic Explorer 1 40km and equation of state held constant} 
\begin{tabular}{|c|c|c|c||c|c|c|c|} 
\hline\multicolumn{4}{|c||}{ $90^{th} \%$ Stiff } & \multicolumn{4}{|c|}{ Maximum Likelihood } \\ \hline 
Mass$_{\mathrm{BH}}$ [$M_{\odot}$] & Distance [Mpc] & SNR & $\ln\mathcal{B}$ & Mass$_{\mathrm{BH}}$ [$M_{\odot}$] & Distance [Mpc] & SNR & $\ln\mathcal{B}$ \\ \hline 5 & 40 & 2888 & 132.3 & 5 & 40 & 2888 & 22.2 \\ \hline 
10 & 40 & 3617 & 7.5 & 10 & 40 & 3617 & 2.1 \\ \hline 
15 & 40 & 4055 & 1.9 & 15 & 40 & 4055 & 0.5 \\ \hline 
20 & 40 & 4390 & 0.3 & 20 & 40 & 4390 & 0.0 \\ \hline 
5 & 80 & 1444 & 30.4 & 5 & 80 & 1444 & 7.1 \\ \hline 
10 & 80 & 1809 & 0.2 & 10 & 80 & 1809 & -0.4 \\ \hline 
15 & 80 & 2028 & 0.8 & 15 & 80 & 2028 & 0.2 \\ \hline 
20 & 80 & 2195 & -0.0 & 20 & 80 & 2195 & 0.1 \\ \hline 
\end{tabular} 
\label{Tab:NSBH_CE1_const}
\end{table} 

\begin{table}[ht] 
\caption{neutron star–black hole with Cosmic Explorer 2 40km} 
\begin{tabular}{|c|c|c|c||c|c|c|c|} 
\hline\multicolumn{4}{|c||}{ $90^{th} \%$ Stiff } & \multicolumn{4}{|c|}{ Maximum Likelihood } \\ \hline 
Mass$_{\mathrm{BH}}$ [$M_{\odot}$] & Distance [Mpc] & SNR & $\ln\mathcal{B}$ & Mass$_{\mathrm{BH}}$ [$M_{\odot}$] & Distance [Mpc] & SNR & $\ln\mathcal{B}$ \\ \hline 5 & 40 & 4385 & 294.1 & 5 & 40 & 4385 & 51.2 \\ \hline 
10 & 40 & 5491 & 16.5 & 10 & 40 & 5491 & 4.2 \\ \hline 
15 & 40 & 6151 & 3.4 & 15 & 40 & 6151 & 1.7 \\ \hline 
20 & 40 & 6655 & 1.1 & 20 & 40 & 6655 & 0.4 \\ \hline 
5 & 80 & 2193 & 70.2 & 5 & 80 & 2193 & 14.6 \\ \hline 
10 & 80 & 2745 & 2.3 & 10 & 80 & 2745 & -0.3 \\ \hline 
15 & 80 & 3075 & 1.1 & 15 & 80 & 3075 & 0.4 \\ \hline 
20 & 80 & 3328 & 0.3 & 20 & 80 & 3328 & 0.2 \\ \hline 
\end{tabular} 
\label{Tab:NSBH_CE2}
\end{table} 

\begin{table}[ht] 
\caption{neutron star–black hole with Cosmic Explorer 2 40km and equation of state held constant} 
\begin{tabular}{|c|c|c|c||c|c|c|c|} 
\hline\multicolumn{4}{|c||}{ $90^{th} \%$ Stiff } & \multicolumn{4}{|c|}{ Maximum Likelihood } \\ \hline 
Mass$_{\mathrm{BH}}$ [$M_{\odot}$] & Distance [Mpc] & SNR & $\ln\mathcal{B}$ & Mass$_{\mathrm{BH}}$ [$M_{\odot}$] & Distance [Mpc] & SNR & $\ln\mathcal{B}$ \\ \hline 5 & 40 & 4385 & 295.6 & 5 & 40 & 4385 & 52.5 \\ \hline 
10 & 40 & 5491 & 16.9 & 10 & 40 & 5491 & 4.4 \\ \hline 
15 & 40 & 6151 & 3.7 & 15 & 40 & 6151 & 1.4 \\ \hline 
20 & 40 & 6655 & 0.8 & 20 & 40 & 6655 & 0.1 \\ \hline 
5 & 80 & 2193 & 71.7 & 5 & 80 & 2193 & 15.8 \\ \hline 
10 & 80 & 2745 & 1.6 & 10 & 80 & 2745 & -0.2 \\ \hline 
15 & 80 & 3075 & 1.3 & 15 & 80 & 3075 & 0.2 \\ \hline 
20 & 80 & 3328 & -0.1 & 20 & 80 & 3328 & -0.1 \\ \hline 
\end{tabular} 
\label{Tab:NSBH_CE2_const}
\end{table} 

\end{document}